\documentclass[prl,twocolumn,showpacs,superscriptaddress,floatfix, reprint]{revtex4}

\usepackage[OT1]{fontenc}
\usepackage[usenames,dvipsnames]{color}
\usepackage[latin1]{inputenc}
\usepackage[english]{babel}
\usepackage{graphicx}
\usepackage{color}
\usepackage{amssymb,amsmath}
\usepackage[Gray,squaren]{SIunits}
\usepackage{xspace}
\usepackage{upgreek}
\usepackage{ulem}
\usepackage{epstopdf}
\newcommand{\ud}{\mathrm{d}}
\newcommand{\vbar}{\bar{v}}
\newcommand{\Iin}{I_0}
\newcommand{\Iout}{I_t}
\newcommand{\Isca}{I_s}
\newcommand{\Isat}{I_{\mathrm{sat}}}
\newcommand{\Ein}{E_0}
\newcommand{\Eout}{E_t}
\newcommand{\Esca}{E_s}
\newcommand{\psca}{\theta_s}
\newcommand{\acos}{\operatorname{acos}}
\newcommand{\ie}{i.e.\@\xspace}

\newcommand{\eq}[1]{Eq.~\eqref{eq:#1}}
\newcommand{\eqs}[2]{Eqs.~\eqref{eq:#1} and~\eqref{eq:#2}}

\newcommand{\fig}[1]{Fig.~\ref{fig:#1}}

\begin{document}

\title{Cooperative Emission of a Coherent Superflash of Light}
\author{C. C. Kwong}
\affiliation{School of Physical and Mathematical Sciences, Nanyang Technological University, 637371 Singapore, Singapore}
\author{T. Yang}
\affiliation{Centre for Quantum Technologies, National University of Singapore, 117543 Singapore, Singapore}
\author{M. S. Pramod}
\affiliation{Centre for Quantum Technologies, National University of Singapore, 117543 Singapore, Singapore}
\author{K. Pandey}
\affiliation{Centre for Quantum Technologies, National University of Singapore, 117543 Singapore, Singapore}
\author{D. Delande}
\affiliation{Laboratoire Kastler Brossel, UPMC-Paris 6, ENS, CNRS; 4 Place Jussieu, 75005 Paris, France}
\author{R. Pierrat}
\affiliation{ESPCI ParisTech, PSL Research University, CNRS, Institut Langevin, 1 rue Jussieu, F-75005, Paris, France}
\author{D. Wilkowski}
\affiliation{School of Physical and Mathematical Sciences, Nanyang Technological University, 637371 Singapore, Singapore}
\affiliation{Centre for Quantum Technologies, National University of Singapore, 117543 Singapore, Singapore}
\affiliation{Institut Non Lin\'eaire de Nice, Universit\'e de Nice Sophia-Antipolis, CNRS UMR 7335, 06560 Valbonne, France}

\date{\today{}}

\begin{abstract}
   We investigate the transient coherent transmission of light through an optically thick cold strontium gas. We observe
   a coherent superflash just after an abrupt probe extinction, with peak intensity more than three times the incident one.
   We show that this coherent superflash is a direct signature of the cooperative forward emission of the
   atoms. By engineering fast transient phenomena on the incident
   field, we give a clear and simple picture of the physical mechanisms at play.
\end{abstract}


\maketitle

\begin{figure}[!htb]
   \centering
   \includegraphics*[width=0.9\linewidth]{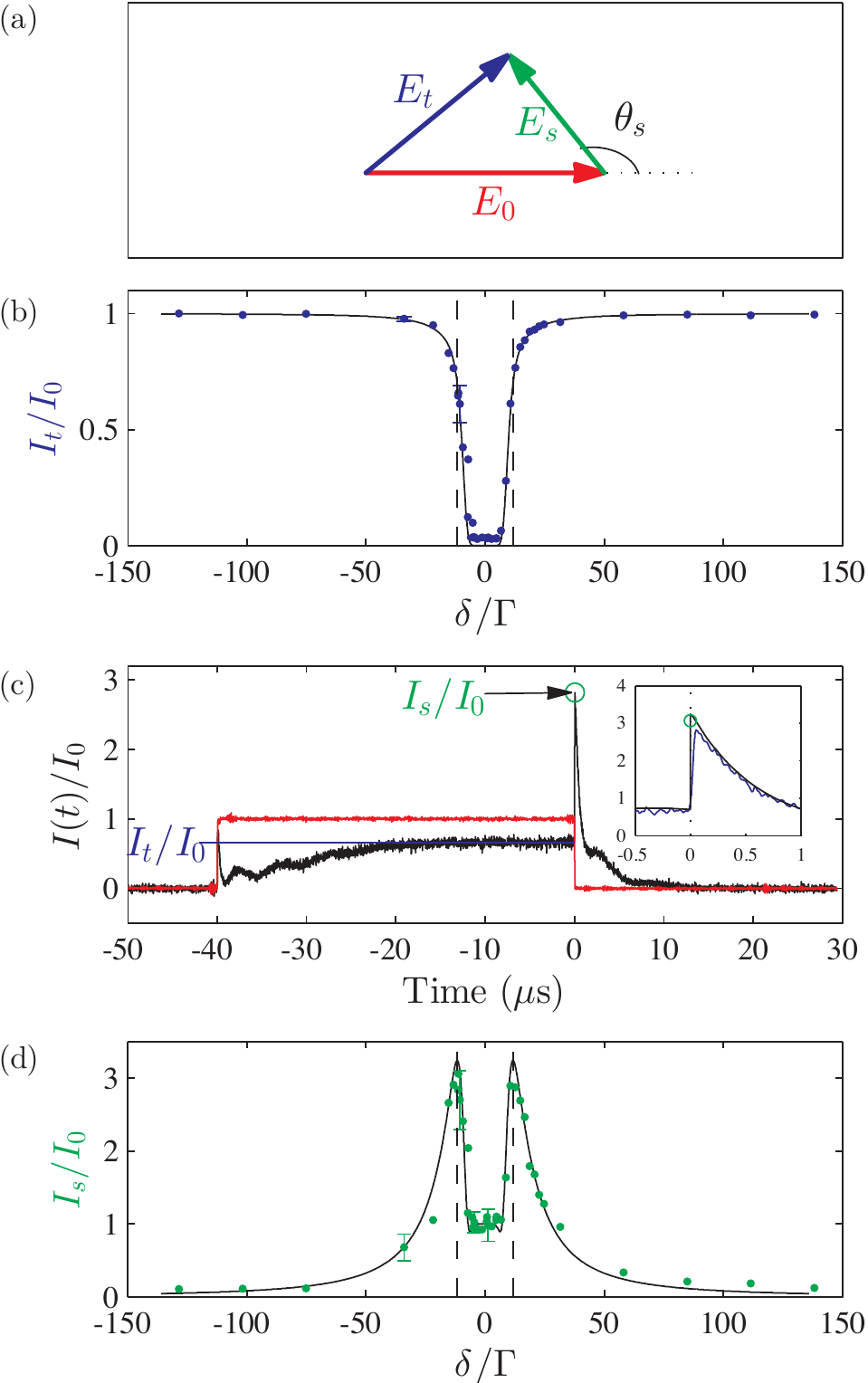}
   \caption{(Color online) (a) A schematic representation of the electric fields in the complex plane. (b) Transmission $\Iout/\Iin$ in the stationary regime as a
   function of the probe detuning $\delta/\Gamma$. The blue dots are the experimental data and the black solid line is
   the theoretical prediction. (c) Temporal evolution of the normalized transmitted intensity for
   $\delta=-11.2\Gamma$. The red curve shows the normalized incident intensity, the black curve the experimental
   signal, the blue line the level of $\Iout/\Iin$ and the green open circle the value of $\Isca/\Iin$. The
   inset is a zoom around $t=0$ of the coherent superflash, with the black curve showing the theoretical prediction
   assuming instantaneous switch off of the probe. (d) $\Isca/\Iin$ in the stationary regime as a function of the probe
   detuning. The black solid line is the theoretical prediction and the green dots are the experimental data. The
   vertical dashed lines at $|\delta| = 11.7\Gamma$ in~(b) and~(d) show the expected positions of the maximum values
   of the forward scattered intensity. In the experiment, $T=\unit{3.3(2)}{\micro\kelvin}$ and $b_0=19(3)$, the other
   parameters being given in the text.}
   \label{fig:stat_reg}
\end{figure}

For many decades, coherent transient phenomena have been used to characterize decays and dephasing in resonantly driven two-level systems \cite{slichter1990principles,allen1974}. A rich variety of systems, with their own particularities, ranging from NMR~\cite{bloembergen1948,hahn1950nuclear} to electromagnetic resonances in atoms \cite{toyoda1997optical,peitilainen1998,zamith2001observation,chalony2011}, molecules \cite{brewer1972optical,foster1974interference,brewer1976optical,judson1992teaching} and nuclei~\cite{helisto1982,ikonen1985}, have been used.
A simple situation arises when an electromagnetic wave is sent through a sample composed of atomic (or molecular) scatterers. The abrupt switch off of a monochromatic quasiresonant excitation leads to free induction decay in the forward direction \cite{brewer1972optical}. Temporal shapes and characteristic decay times of free induction decay depend on quantities such as laser frequency detuning \cite{toyoda1997optical}, optical thickness \cite{shim2002optical,chalony2011}, and on the presence of inhomogeneous broadening \cite{brewer1972optical} and nonlinearities \cite{ducloy1977}. For an optically thick medium, since the incoming light is almost completely depleted by scattering in the stationary regime, the free induction decay signal takes the form of a coherent flash of light \cite{chalony2011}. Its duration is reduced with respect to the single scatterer lifetime by a factor of the order of the optical thickness \cite{chalony2011}. Consequently, its experimental observation, using standard optical
transitions (lifetime in the nanosecond range), is rather challenging \cite{wei2009}.
In this Letter, we solve this issue by performing free induction decay on the intercombination line of a cold strontium atomic gas. We gain physical insight into coherent transmission, and observe a coherent superflash of light, i.e., a transmitted
intensity larger than the incident one~[see \fig{stat_reg}(c)]. The superflash is due to strong phase rotation and large amplitude of the forward scattered field
which are directly measured in our experiment.

Related effects have been observed in M\"ossbauer spectroscopy experiments, where a temporal phase change in the $\gamma$ radiation can lead to transient
oscillations of the intensity transmitted through a sample~\cite{helisto1982,ikonen1985}.
These oscillations are rather small, typically of the order of 1\%.
This is because the $\gamma$ emitter used has a short coherence time. Note that no superflash
was ever observed.
In a refined ``$\gamma$ echo'' experiment, a coincidence detection made it possible
to shift the phase of the emitter at a specific time during its exponential
decay, leading to a revival of the forward transmitted intensity~\cite{helisto1991}.
Laser spectroscopy is, however, a much easier and flexible tool. First, the
temporal or spectral properties of the source can be tuned almost at will, and second, a dilute cold atomic gas
can be thought of as a collection of independent identical highly-resonant two-level systems.

We first consider a scheme where a laser beam is sent through a slab uniformly filled with resonant pointlike
scatterers. In the stationary regime, scattering leads to an attenuation of the intensity, $\Iout=|\Eout|^2$, of the
transmitted coherent field $\Eout$, according to the Beer-Lambert law
\begin{equation}
   \Iout=\Iin\exp\left(-b\right),
   \label{eq:lb_law}
\end{equation}
where $\Iin=|\Ein|^2$ is the intensity of the incident field $\Ein$ and $b$ is the optical thickness of the medium. The power
lost in the coherent transmission, $\propto 1-e^{-b}$, leaves the medium in all
directions~\cite{fioretti1998observation,labeyrie2003slow}. In general, since the positions of the scatterers are
random, the reemitted field is incoherent (\ie, the phase of the incident field is lost). This statement is, however,
not true in the forward direction, where the phase of the scattered field does not depend on the (transverse) positions
of the scatterers~\footnote{Partial coherence exists in the backward direction as well. See, for example, ~\cite{kaiser2002} and
references therein.}.
This cooperative effect of the atomic ensemble in the forward direction has already been
explored by several authors, for example in superradiance laser~\cite{meiser2009,bohnet2012}, superradiance of a single
photon emission~\cite{scully2009}, and in the underlying mechanical effects on the atomic cloud~\cite{bienaime2010}.
Importantly, the attenuation of the transmitted field can be interpreted as the result of a destructive interference
between the incident field and the field scattered in the forward direction.  Denoting the forward scattered
electric field by $\Esca$, one has at any time
\begin{equation}
   \Eout=\Ein+\Esca.
   \label{eq:field_relation}
\end{equation}
For the useful case of a monochromatic field at frequency $\omega$ in the stationary regime, such an equality can be
written for the complex field amplitudes. A geometrical representation of~\eq{field_relation} is shown
in~\fig{stat_reg}(a), where the angle $\psca$ represents the relative phase between $\Esca$ and $\Ein$. In general, the
fields have two polarization components, so vectors should be used. Here,
we consider a simpler situation where all fields have the same polarization.

In the stationary regime, energy conservation imposes $\Iout\leq\Iin$. In other words, $|\Eout| \leq
|\Ein|$, and therefore, $|\Esca|\leq 2|\Ein|$. However, since the forward scattered field is built upon the incident
field, one might believe that its amplitude is bounded as such, $|\Esca|\leq|\Ein|$. As a key result of this Letter, we
show that the latter intuitive picture is incorrect. Indeed, we predict a forward scattered intensity $\Isca$ arbitrarily
close to $4\Iin$ and experimentally observe $\Isca/\Iin=3.1$. The experimental value is mainly limited by
the maximum optical thickness that can be obtained with our experimental setup. Hence, apart
from the energy conservation argument, we find no other basic principles or theorems, such as causality or
Kramers-Kronig relations, that limit the amplitude of the forward scattered field.

The system under investigation consists of a laser-cooled \textsuperscript{88}Sr atomic gas. The details of the cold
atoms production line are given in Ref.~\cite{tao2014}. The last cooling stage is performed on the
$^1S_0\,\rightarrow\,^3P_1$ intercombination line at transition wavelength $\lambda=\unit{689}{\nano\meter}$, with
a bare linewidth of $\Gamma/2\pi=\unit{7.5}{\kilo\hertz}$. The number of atoms is $2.5(5)\times 10^8$. The temperature
of the cold gas is $T=\unit{3.3(2)}{\micro\kelvin}$, corresponding to an rms velocity of $\vbar=3.4\Gamma/k$.
Here, $k=2\pi/\lambda$ is the wave vector of the transition. The
cloud has an oblate ellipsoidal shape with an axial radius $\unit{240(10)}{\micro\meter}$ and an equatorial radius
$\unit{380(30)}{\micro\meter}$ with a peak density around $\rho=\unit{4.6\times 10^{11}}{\rpcubic{\centi\meter}}$.  Using shadow imaging technique, we measure along an equatorial direction of the cloud,
an optical thickness at resonance of $b_0=19(3)$. We note that $k\ell\simeq 500$, where
$\ell$ is the light scattering mean free path. Since $k\ell\gg 1$, the system is deeply in the dilute regime. Hence, all
collective behaviors in dense media such as Dicke superradiance in free space~\cite{dicke1954,rehler1971superradiance,bonifacio1975cooperative,gross1982superradiance}, recurrent scattering~\cite{sokolov2009,changchi2014}, Lorentz-Lorenz and collective Lamb
shift~\cite{pellegrino2014,keaveney2012,javanainen2014}, can be disregarded. Atomic collisions are also negligible
over the duration of the experiment in our dilute cold gas.

A probe laser beam is then sent across the cold atomic gas along an equatorial axis. The probe (diameter
$\unit{150}{\micro\meter}$) is tuned around the resonance of the intercombination line. Its power is
$\unit{400(40)}{\pico\watt}$, corresponding to $0.45(5)\Isat$, where
$\Isat=\unit{3}{\micro\watt\per\centi\meter\squared}$ is the saturation intensity of the transition. The probe is
switched on for $\unit{40}{\micro\second}$ such that the stationary regime is reached without introducing significant
radiation pressure on the atoms. The same probe sequence is repeated $\unit{1}{\milli\second}$ later without the atoms
to measure $\Iin$.
The transmitted photons along the propagation direction are
collected on a photodetector, leading to a transverse integration of the intensity. During probing, we apply a bias magnetic field of
$\unit{1.4}{G}$ along the beam polarization to address a two-level system corresponding to the $^1S_0,
m=0\,\rightarrow\,^3P_1, m=0$ transition.

First, we look at the stationary regime. We plot, in~\fig{stat_reg}(b), $\Iout/\Iin$ as a function of the probe frequency
detuning $\delta$. To compare with analytical predictions, we model the ellipsoid geometry of the cloud by a slab
geometry. In the frequency domain, the coherent transmitted electric field through the slab is given by
\begin{equation}
   \Eout(\omega)=\Ein(\omega)\exp\left[i\frac{n(\omega)\omega L}{2c}\right].
   \label{eq:field_trans}
\end{equation}
We define, $n(\omega)$, $\omega$, $c$, and $L$, respectively, as the complex refractive index, the laser optical
frequency, the speed of light in vacuum, and the thickness of the slab along the laser beam. For a dilute medium, we have
$n(\omega) = 1 + \rho\alpha(\omega)/2$~\cite{Hetch1974}. The two-level atomic polarizability is given by
\begin{equation}
   \alpha(\omega)=-\frac{3\pi\Gamma c^3}{\omega^3}\frac{1}{\sqrt{2\pi}\vbar}\int^{+\infty}_{-\infty}\ud v
      \frac{\textrm{exp}\left(-v^2/2\vbar^2\right)}{\delta-kv+i\Gamma/2},
   \label{eq:polarizability}
\end{equation}
where the integration is carried out over the thermal Gaussian distribution of the atomic velocity $v$ along the beam
propagation direction (Doppler broadening). By inserting, in~\eq{field_trans}, the polarizability, the measured
values of the atomic density, and the temperature, we compute the transmitted intensity $\Iout$ and show the results
in~\fig{stat_reg}(b). The effective slab thickness of the cloud is chosen to match the measured optical thickness.  The
theoretical prediction agrees very well with the experimental data.  However, close to resonance, the measured
transmission is slightly higher than predicted. This mismatch is due to the finite transverse size of the cloud,
which allows few photons in the wings of the laser beam to be directly transmitted.

We now take advantage of the finite response time of the light-atom system to measure the forward scattered
intensity directly. For this purpose, we abruptly switch off the probe beam. The switching time is $\unit{40}{\nano\second}$ (\ie
$\sim 500$ times faster than the excited state lifetime $\Gamma^{-1}=\unit{21}{\micro\second}$). According
to~\eq{field_relation}, if $\Iin=0$, we have $\Eout(t=0^+)=\Esca$. Hence, immediately after switching off the probe, the
detector measures the forward scattered intensity of the stationary regime [\ie, $\Iout(t=0^+)=\Isca$]. In the absence of a driving field, free induction decay occurs.
If the probe is at
resonance and the optical thickness is large, the stationary transmitted intensity is very small, i.e., $\Eout(t=0^-)\simeq 0,$
so that $\Esca(t=0^-)\simeq - \Ein.$ Immediately after the probe is switched off, the atomic field $\Esca$ does not change, so that
$\Eout(t=0^+)\simeq -\Ein$ and $\Iout=\Iin$.
The free induction decay, thus, leads to the emission of a coherent flash of
light with a peak intensity equal to $\Iin$ (see for example Fig.~1(b) in Ref.~\cite{chalony2011}). For a detuned probe
field, one illustrative example of the temporal evolution of $\Iout/\Iin$ is given in~\fig{stat_reg}(c). In this case,
we observe a flash of light with the peak intensity clearly above $\Iin$. We define it as a \emph{coherent superflash}.
In the
inset of~\fig{stat_reg}(c), we compare the experimental signal and the theoretical prediction. This theoretical
prediction is obtained by numerically calculating the inverse Fourier transform of \eq{field_trans} for an
incident field that is a step function in the time domain. A good agreement is obtained, except at $t=0$, where the
finite response time of our detection scheme slightly smoothes the predicted discontinuity. The value of $\Isca$ is
obtained by extrapolating the (super)flash down to $t=0$.

We plot in~\fig{stat_reg}(d) the normalized forward scattered intensity as a function of the
laser detuning. At resonance, we find $\Isca/\Iin\simeq 1$ and $\Iout/\Iin\simeq 0,$ as in Ref.~\cite{chalony2011},
meaning that the interference between the incident field and the forward scattered field is almost perfectly destructive
(\ie $\Esca\simeq-\Ein$). Far from resonance, $\Isca\overrightarrow{_{|\delta|\rightarrow \infty}} 0$ so that
$\Iout\overrightarrow{_{|\delta|\rightarrow \infty}} \Iin.$ In between these two extreme cases, $\Isca/\Iin$ passes
through a maximum of $3.1(4)$ at $|\delta|=11.2(7)\Gamma$. At the same detuning, we get
$\Iout/\Iin=0.66(8)$. Finding $\Isca>\Iin$ (\ie a coherent superflash) is surprising for two reasons. First, as
mentioned earlier, the forward scattered field is built upon the incident field. Second, it reaches its maximum value
when the field is mostly transmitted (\ie where we could expect that the incident field weakly interacts with the
medium).

\begin{figure}[!htb]
   \centering
   \includegraphics*[width=0.9\linewidth]{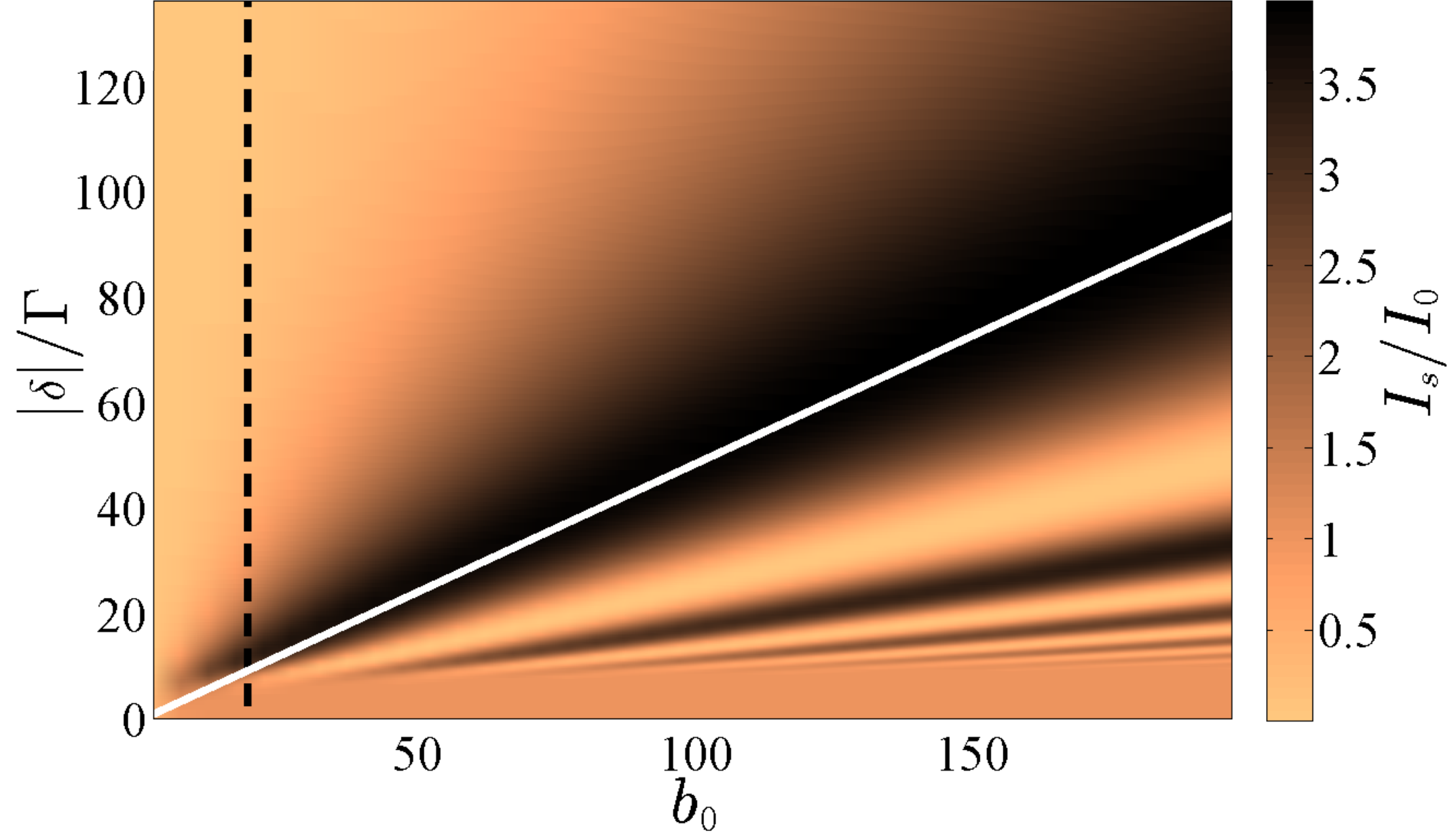}
   \caption{(Color online) Prediction for the $\Isca/\Iin$ ratio vs. parameters $b_0$ (optical thickness at resonance)
   and detuning $|\delta|/\Gamma$ for $T=\unit{3.3(2)}{\micro\kelvin}$. The black dashed line indicates the optical
   thickness of our experiment. The white solid line represents the linear dependence on $b_0$ of the detuning at
   which maximum value of $\Isca/\Iin$ is attained. }
   \label{fig:super_flash}
\end{figure}

\begin{figure}[!htb]
   \centering
   \includegraphics*[width=0.8\linewidth]{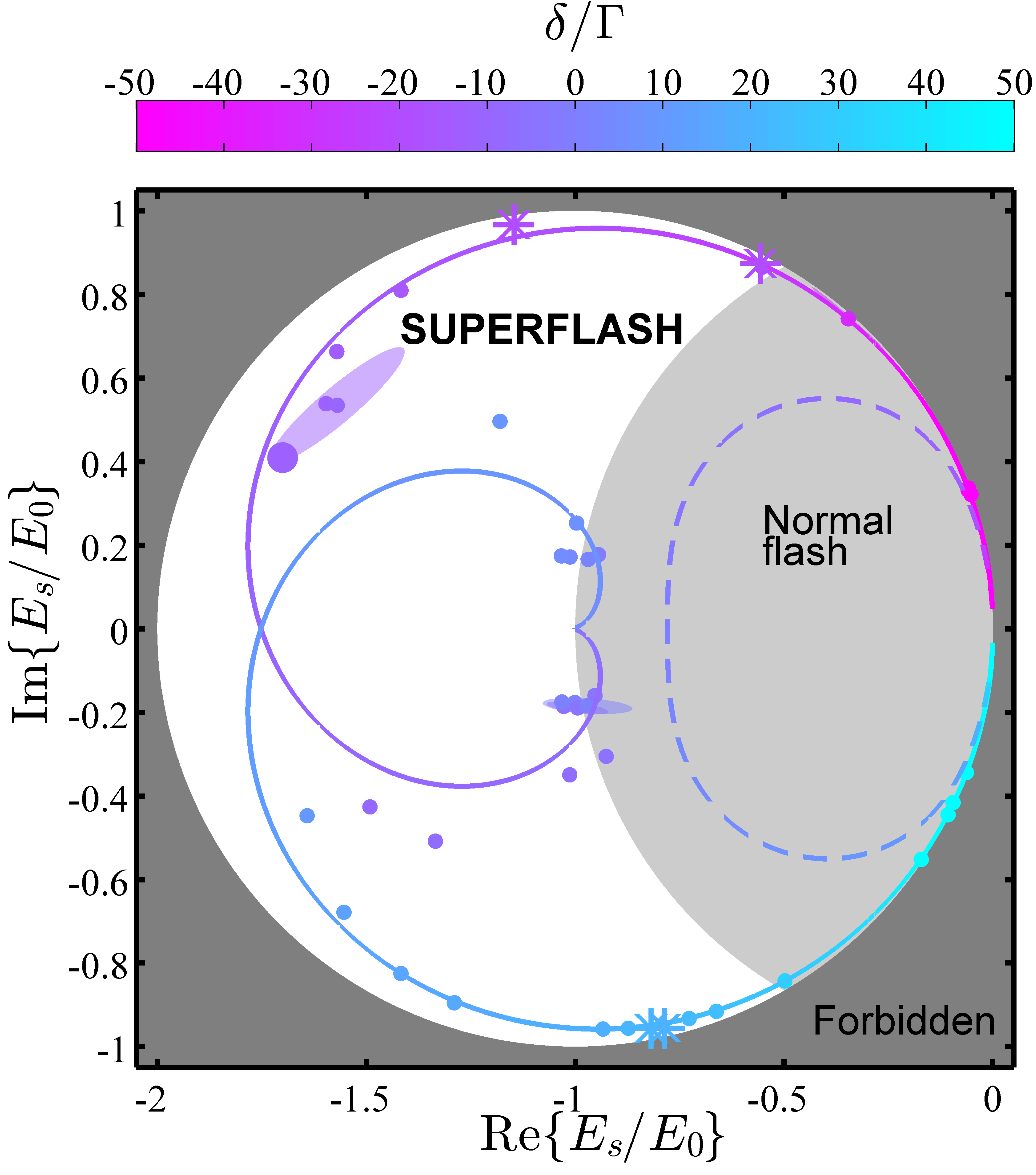}
   \caption{(Color online) Reconstruction of $\Esca$ in the complex plane. The false color scale gives the probe
   detuning. The white region corresponds to the superflash regime, the light gray area shows the region with the normal
   coherent flash, and the dark gray area gives the region forbidden by energy conservation. The dots and stars are the
   experimental values (see text for more details). The transparent ellipses around several experimental data depict the error
   estimates. The solid and dashed curves are theoretical predictions, respectively, at $b_0=19$ and $b_0=3$ for a temperature $T=\unit{3.3}{\micro\kelvin}$.}
   \label{fig:phase}
\end{figure}

In the stationary regime, energy conservation imposes the transmitted intensity to be lower than the incident one, which, from \eq{field_relation},
implies $|\Ein+\Esca|^2\leq |\Ein|^2.$ Thus, $\Esca$ must lie inside a circle (represented by the white and light grey areas in \fig{phase})
with center $-\Ein$ and radius $|\Ein|.$
The maximum $|\Esca|$ is, thus, reached when $\Esca=-2\Ein,$ implying that the maximum superflash is $\Isca=4\Iin.$  We tend to this limit as we
increase the optical thickness, as shown in~\fig{super_flash}. For $|\delta| \gg k\bar{v}$, the maximum superflash
intensity at a given large $b_0$ occurs for $\psca\approx\pi$.  This
corresponds to $|\delta| / \Gamma \approx b_0 / 4\pi g(k\bar{v}/\Gamma)$ where $g(x) =
\sqrt{\pi/8}\exp(1/8x^2)\operatorname{erfc}(1/\sqrt{8}x)/x$~\cite{chalony2011}. At this detuning, the superflash intensity is $I_s/I_0
\approx 4[1-2\pi^2g(k\bar{v}/\Gamma)/b_0]$. At $T = \unit{3.3}{\micro\kelvin}$, the temperature of the experiment,
$g(k\bar{v}/\Gamma) = 0.16$.  The detuning at maximum superflash intensity is then given by $|\delta|/\Gamma \approx
0.48 b_0$, a linear dependence on $b_0$ which can be seen in~\fig{super_flash}.

From our experimental measurements of  $\Iout/\Iin$ (stationary transmitted probe intensity) and $\Isca/\Iin$
(immediately after switching off the probe), we extract the phase of the forward scattered field
\begin{equation}
   \theta_{s}=\acos\left(\frac{\Iout-\Iin-\Isca}{2\sqrt{\Iin\Isca}}\right).
   \label{eq:angle}
\end{equation}
However, an ambiguity exists in the phase calculated using~\eq{angle}, since we cannot distinguish between $\theta_s$
and $-\theta_s$. To disambiguate, the easiest way is to choose the sign giving the best agreement
with~\eqs{field_trans}{polarizability}. The result of this procedure is represented by the dots in~\fig{phase}.

We have also added theoretical predictions in~\fig{phase}. We note that the phase angle $\psca$ is within the range
$[\pi/2,3\pi /2]$ (see the allowed circle in \fig{phase}), which means that the forward scattered field always destructively interferes with $\Ein$,
a necessary condition for a passive scattering medium. We also note that for large detunings $\psca\overrightarrow{_{\delta\rightarrow \pm\infty}}\mp\pi/2$. However, $|\Esca|$ is close to zero, so $\Esca$ stays close to the origin.  As the detuning decreases, $\Esca$
traces a curve as depicted in~\fig{phase} until we reach a situation where $\psca\simeq\pi$. If this happens when the
detuning is still relatively large, as it is in the experiment, a large superflash intensity is observed.

At very large optical thickness, $\psca$ goes back and forth in the $[\pi/2,3\pi /2]$ range, leading to a potential
observation of several superflashes by scanning the detuning at a given $b_0,$ see ~\fig{super_flash}. At low optical
thickness, the excursion of $\psca$ is limited and no superflash occurs as is illustrated by the dashed
curve in~\fig{phase}.

\begin{figure}[!htb]
   \centering
   \includegraphics*[width=0.9\linewidth]{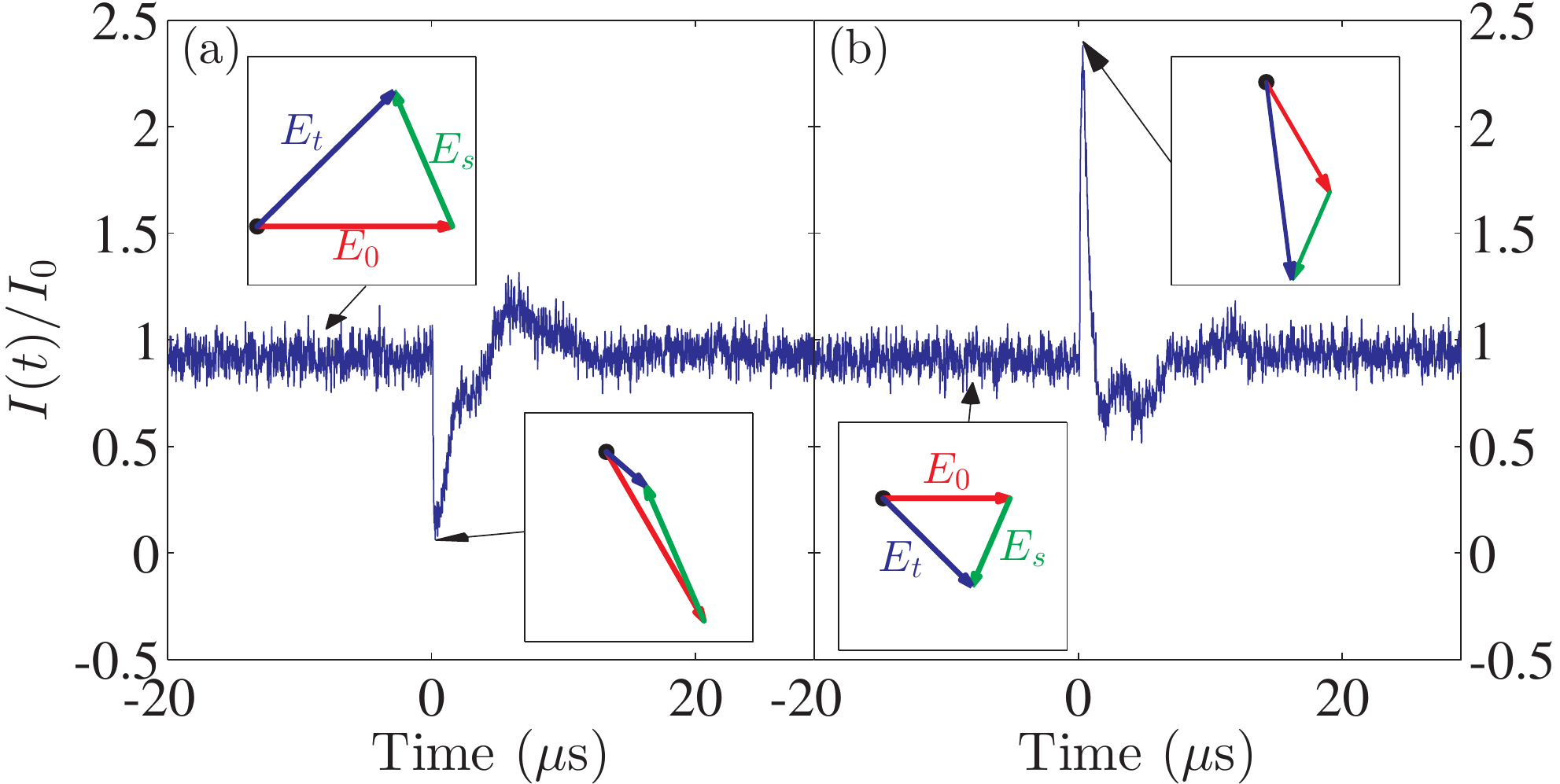}
   \caption{(Color online) Temporal evolution (blue curves) with an abrupt change of phase of $-0.4\pi$ at $t=0$ for a
   probe detuning of (a) $\delta=-19.3\Gamma$ and (b) $\delta=+20.7\Gamma$. The insets show a schematic representation of the electric fields in the
   complex plane at the time pointed by the arrows. }
   \label{fig:phase_recons}
\end{figure}

An additional measurement makes it possible to disambiguate the sign of phase $\psca$.  We insert, in the optical
path of the probe, an electro-optic modulator (EOM) to adjust the phase delay of $\Ein$.
By abruptly switching off
the EOM bias voltage, we create an abrupt negative jump in the phase of $\Ein$.  Depending on whether $\Ein$ interferes
constructively or destructively with $\Esca$ after the phase jump, we observed a positive (super)flash
[see~\fig{phase_recons}(b)] or a negative flash [see~\fig{phase_recons}(a)], respectively.
We further vary the phase jumps in
the $[0,-\pi]$ range where the amplitude of the (super)flash necessarily passes through an extremum giving, without
ambiguity, $\psca$. We show as stars in~\fig{phase} several values such a reconstructed field.

In conclusion, we have studied fast transient phenomena in the transmission of a probe beam through an optically thick
cold atomic sample.  When a detuned probe is abruptly
switched off, a short coherent superflash is emitted with a peak intensity up to 4 times the incident intensity. By
combining transient and stationary intensity measurements, we show that the coherent superflash comes
from a phase rotation of the forward scattered field induced by the large optical thickness of the medium.
The sensitivity of the transmitted intensity to the changes in the phase of the incident field suggests
that an optically dense medium may be useful as a phase discriminator device and as a generator of pulse trains with repetition rates higher than $\Gamma$ \cite{changchipulse2014}.

The authors are grateful to R. Carminati, C. Salomon and J. Ye for fruitful discussions and to a referee
for his-her very valuable comments. C. C. K. thanks the CQT and ESPCI
institutions for funding his trip to Paris.  This work was supported by the CQT/MoE funding Grant No. R-710-002-016-271.
R. P. acknowledges the support of LABEX WIFI (Laboratory of Excellence Grant No. ANR-10-LABX-24) within the French Program
``Investments for the Future'' under Reference No. ANR-10-IDEX-0001-02 PSL$^{\ast}$.

\bibliographystyle{apsrev}

\newcommand{\noopsort}[1]{}\providecommand{\noopsort}[1]{}\providecommand{\singleletter}[1]{#1}%

\end{document}